\documentclass[%
floatfix,
aps,
 jor,
 amsmath,amssymb,
 reprint,%
]{revtex4-2}

\usepackage{graphicx}% Include figure files
\usepackage{dcolumn}% Align table columns on decimal point
\usepackage{bm}% bold math
\usepackage{amsmath}
\usepackage{subfig}
\usepackage{floatrow}
\usepackage{hyperref}
\usepackage[autostyle]{csquotes}
\usepackage{microtype}
\usepackage{comment}

\hypersetup{
    colorlinks=true,
    linkcolor=blue,
    filecolor=magenta,      
    urlcolor=cyan,
    pdftitle={Overleaf Example},
    pdfpagemode=FullScreen,
    }
    
\begin{document}

\preprint{AIP/123-QED}

\title[Sample title]{\textbf{Spontaneous collective transport in a heat--bath}
%\footnote{Error!}
}% Force line breaks with \\
%\thanks{Footnote to title of article.}

\author{Mayank Sharma}
 \email{mayank.sharma@students.iiserpune.ac.in}
 %\altaffiliation[Also at ]{Physics Department, XYZ University.}%Lines break automatically or can be forced with \\
\author{A. Bhattacharyay}%
 \email{a.bhattacharyay@iiserpune.ac.in}
\affiliation{ 
Indian Institute of Science Education and Research, Pune, India%\\This line break forced with \textbackslash\textbackslash
}%

%\author{C. Author}
 %\homepage{http://www.Second.institution.edu/~Charlie.Author.}
%\affiliation{%
%Second institution and/or address%\\This line break forced% with \\
%}%

\date{\today}% It is always \today, today,
             %  but any date may be explicitly specified

\begin{abstract}
We investigate emergence of spontaneous filtering of Brownian motion in higher dimensional space by many--body structures of symmetry broken dimer. Interacting dimer form structures which eventually restrict rotational degrees of freedom of individual constituents resulting in spontaneous emergence of collective transport. In this phenomenon, interaction and broken structural symmetry play cooperatively in emergence of collective transport out of thermal fluctuations. As a general phenomenon, spontaneous filtering of Brownian fluctuation might play an important role in the structural transition of complex molecules on top of what is known in terms of Kramers--rate process. This could also be a missing link to completely understand basics of bio--polymers' folding transitions (protein folding) where structure might encode motion in phase space.
\end{abstract}

%Use showkeys class option if keyword
                              %display desired
\maketitle

 %\begin{quotation}
% The ``lead paragraph'' is encapsulated with the \LaTeX\ 
% \verb+quotation+ environment and is formatted as a single paragraph before the first section heading. 
% (The \verb+quotation+ environment reverts to its usual meaning after the first sectioning command.) 
% Note that numbered references are allowed in the lead paragraph.
% %
% The lead paragraph will only be found in an article being prepared for the journal \textit{Chaos}.
% \end{quotation}

 %\section{\label{sec:level1} Main
% }
\onecolumngrid
Coordinate dependence of damping/diffusion arises near a boundary or interface due to presence of slow hydrodynamic modes generated by thermal fluctuations of a Brownian particle \cite{Faxen1924, Cox1967TheSM, Brenner1961TheSM}. Existence of this coordinate/state dependence of damping/diffusion entails new physics for structured mesoscopic objects under thermal equilibrium conditions. The most striking emergent consequence of this is that, average response of the structured mesoscopic object to homogeneous and isotropic fluctuations of the heat bath could be be symmetry broken \cite{bhattacharyay2012directed,sharma2020conversion}. This should not be any surprise in the presence of additional slow hydrodynamic modes generated by Brownian fluctuations themselves which on comparable time scales get renormalized by these modes. The gross effect of the presence of these slow modes being renormalization of damping/diffusivity, when taken into account within the framework of Brownian motion, it can spontaneous filter Brownian motions of a structured mesoscopic object in a homogeneous and isotropic heat-bath \cite{bhattacharyay2012directed,sharma2020conversion}. The present paper demonstrates collective transport arising in a many-body system of such objects driven solely by thermal fluctuations.
\par
Collective transport like flocking of birds, shoaling of fishes etc., are generally studied by the introduction of a non-equilibrium \enquote{self-driving} force in the context of active systems. One of the earliest models that displays flocking from an initially disordered configuration is the Vicsek model \cite{vicsek1995novel}. There are various variants of the Vicsek model present in literature \cite{chate2008modeling,miguel2018effects,giardina2008collective,cavagna2015flocking}. Many models of active systems have been developed to demonstrate flocking transitions, pattern formation, swarming transitions \cite{liao2020dynamical,martin2018collective, lobaskin2013collective}, motility-induced phase separation (MIPS) \cite{fily2012athermal,redner2013structure,bialke2012crystallization,buttinoni2013dynamical}. 
Redner et al., has shown that an active colloidal system in thermal noise and excluded volume interaction displays phase separation into dense and dilute phases \cite{redner2013structure}. Martin et al., considered an assembly of active Brownian particles with excluded volume interaction and velocity aligning protocol shows emergence of flocking \cite{martin2018collective}. Liao et al., showed emergence of flocking in an assembly of active Brownian disks with steric-repulsion and dipole-dipole interactions \cite{liao2020dynamical}.
\par
In biological systems, matching of molecular structures is of profound importance for functionality. In the stochastic dynamics of complex molecules as well as in various other systems, coordinate/state dependence of diffusion is identified these days to be a key ingredient \cite{roussel2004reaction,barik2005quantum,sargsyan2007coordinate,chahine2007configuration,best2010coordinate,lai2014exploring,berezhkovskii2017communication,foster2018probing,ghysels2017position,yamilov2014position,faucheux1994confined}. When coordinate or state dependence of diffusion can result in directed transport of structured objects by filtering Brownian motion, its influence on barrier overcoming transitions could potentially be immense. Thermal fluctuations driven transport can play its kinematic role in the collective functioning of objects undergoing structural transformations. However, this fundamental possibility has been ignored for long because of the notion of impossibility of having ratcheting without a non-equilibrium drive.
\par
We numerically explore characteristic motion of a collection of interacting symmetry broken dimer (dipolar objects) which undergo Brownian fluctuations in the presence of coordinate dependent damping. Dimers interact by dipole-interaction where similar poles repel and opposite poles attract keeping intact the excluded volume. In one--dimensional space, a symmetry broken dimer can show directed motion caused by coordinate dependence of diffusion \cite{bhattacharyay2012directed, sharma2020conversion}. However, rotational degrees of freedom are present in higher dimensional space and that would average out axial broken symmetry. Structures restrict degrees of freedom of its constituents, therefore, such dimers in many--body structures can have suppressed rotational degrees of freedom at the same time when they have a tendency to move along their axis. Depending upon interactions, if many-body structures formed are such that the axes and direction of translation of individual entities are aligned, collective transport can result. This is the motivation of the present numerical exploration that presents collective transport qualitatively akin to those of active systems in the absence of any non-equilibrium drive.

\section{The model}
\subsection*{Symmetry broken dimer}
Consider over--damped equations of motion of a system of two particles at positions $x_1$ and $x_2$ ($x_1 > x_2$) in a one-dimensional space as
\begin{eqnarray}\nonumber
    \frac{dx_1}{dt} &=& -\frac{1}{\Gamma_1(z)}\frac{\partial V(z)}{\partial x_1}+\sqrt{\frac{2k_BT}{\Gamma_1(z)}}\eta_1(t).\\
    \frac{dx_2}{dt} &=& -\frac{1}{\Gamma_2(z)}\frac{\partial V(z)}{\partial x_2}+\sqrt{\frac{2k_BT}{\Gamma_2(z)}}\eta_2(t).
\end{eqnarray}
where $z=x_1-x_2$ is configuration coordinate, $\eta_i(t)$ is Gaussian white noise of unit strength with zero mean and no cross-correlation i.e. $\langle \eta_i(t_1)\eta_j(t_2) \rangle = \delta_{ij}\delta(t_1-t_2)$, $T$ is temperature of the heat-bath and $k_B$ is the Boltzmann constant. Particles are characterized by different damping coefficients $\Gamma_i(z)$ ($i=1, 2$) which depend on the configuration coordinate $z$ of the system. $V(z)$ generates a conservative force field between these particles which is attractive at a larger distance and repulsive at small separations to account for an excluded volume interaction. Considering the centre mass (CM) of the dimer to be $x = \frac{x_1+x_2}{2}$, one can rewrite the model as
\begin{eqnarray}\nonumber
    \frac{dz}{dt} &=& -\left [ \frac{1}{\Gamma_1(z)}+\frac{1}{\Gamma_2(z)}\right ]\frac{\partial V(z)}{\partial z} + \xi_z(z,t).\\
    \frac{dx}{dt} &=& -\frac{1}{2}\left [ \frac{1}{\Gamma_1(z)}-\frac{1}{\Gamma_2(z)} \right]\frac{\partial V(z)}{\partial z} + \xi_x(z,t).
\end{eqnarray}
where $\xi_z(z,t)=\sqrt{2k_BT}\left [ \frac{\eta_1(t)}{\sqrt{\Gamma_1(z)}} -\frac{\eta_2(t)}{\sqrt{\Gamma_2(z)}} \right ]
$ and $\xi_x(z,t) = \sqrt{k_BT}\left [ \frac{\eta_1(t)}{\sqrt{2\Gamma_1(z)}} +\frac{\eta_2(t)}{\sqrt{2\Gamma_2(z)}} \right ]$, such that $\langle \xi_z(z,t_1)\xi_z(z,t_2)\rangle=2k_BT\left( \frac{1}{\Gamma_1(z)}+\frac{1}{\Gamma_2(z)}\right)\delta(t_1-t_2)$ and $\langle \xi_x(z,t_1)\xi_x(z,t_2)\rangle=\frac{k_BT}{2}\left( \frac{1}{\Gamma_1(z)}+\frac{1}{\Gamma_2(z)}\right)\delta(t_1-t_2)$. In eqn.(2), the configuration space (z-space) has decoupled from that of CM of the system. The average velocity of CM of the system is
\begin{equation}
    \left\langle \frac{dx}{d t}\right\rangle = -\frac{1}{2}\left\langle \left [ \frac{1}{\Gamma_1(z)}-\frac{1}{\Gamma_2(z)} \right]\frac{\partial V(z)}{\partial z} \right\rangle,
\end{equation}
and the average velocity of the configuration coordinate $z$ is
\begin{equation}
    \left\langle \frac{dz}{d t}\right\rangle = -\left\langle \left [ \frac{1}{\Gamma_1(z)}+\frac{1}{\Gamma_2(z)} \right]\frac{\partial V(z)}{\partial z} \right\rangle.
\end{equation}
\par
The equilibrium distribution of the configuration coordinate, according to a standard It\^o--process being
\begin{equation}
    P(z)=\frac{A}{k_BT\left( \frac{1}{\Gamma_1(z)}+\frac{1}{\Gamma_2(z)}\right)} e^{-V(z)/k_BT},
\end{equation}
where $A$ is a normalization constant, we get $\left\langle \frac{dz}{d t}\right\rangle \equiv 0$, however $\left\langle \frac{dx}{d t}\right\rangle \neq 0$ in general for $\Gamma_1(z)\neq\Gamma_2(z)$. The relation $\left\langle \frac{dz}{d t}\right\rangle \equiv 0$ follows from the fact that $\left\langle \frac{dz}{d t}\right\rangle = \int{\frac{dP(z)}{dz}dz} =\int_0^0{d[P(z)]}\equiv 0$, where $P(z)$ vanishes at $V(z)\to\infty$. This result is generally true for any confining potential and the equilibrium distribution of configuration remains stationary despite the CM moving, on average, uniformly through the heat bath.
\par
The term $\left [ \frac{1}{\Gamma_1(z)}-\frac{1}{\Gamma_2(z)} \right]$ being the symmetry breaking term in presence of coordinate dependence of damping \cite{bhattacharyay2012directed,sharma2020conversion} when equilibrium of the nonlinear stochastic system is identified as an It\^o--process, average directed transport of such symmetry broken system through the uniform heat--bath automatically results. This is a very important point to understand that constant damping $\Gamma_1\neq\Gamma_2$ can not break the symmetry of the system resulting in average transport of CM because, in that case, one can write $\left\langle\frac{\partial V(z)}{\partial z}\right\rangle =\int_0^0{dP(z)\equiv 0}$ where $P(z)$ is the equilibrium distribution of configuration $z$ which would be Boltzmann-distribution for coordinate independent damping. Coordinate dependence of damping/diffusion is, therefore, essential ingredient for such transport and this, is the most significant effect of the It\^o--distribution as the equilibrium distribution of such nonlinear stochastic systems \cite{bhattacharyay2019equilibrium, bhattacharyay2020generalization, maniar2021random, dhawan2022distribution}. 

\par
It is also important to note that, even when $\left\langle \frac{dx}{dt}\right\rangle \neq 0$, the other average velocity $\left\langle \frac{dz}{dt}\right\rangle$ has to be identically zero under equilibrium conditions \cite{bhattacharyay2012directed,sharma2020conversion}. Equilibrium fluctuations of the system are function of configuration coordinate $z$. There cannot exist any average current in configuration space because there exists force and average current will produce entropy violating equilibrium condition of detailed balance. Vanishing of $\left\langle \frac{dz}{dt}\right\rangle$, in general, for any choice of $\Gamma_{i}(z)$ and $V(z)$ ensures thermal equilibrium for an It\^o-distribution.

\par
Coordinate dependence of damping will arise, in general, due to hydrodynamics induced by configuration fluctuations of the dimer. However, these hydrodynamic modes, getting generated due to thermal fluctuation of the dimer, do not represent any independent source of energy than bath fluctuations. Note that, the additional inhomogeneity of space induced by coordinate-dependence of damping is taken into account in the equilibrium distribution of an It\^o-process which is in general missed by Boltzmann distribution.
 
 %\section{\label{sec:level II} Many Dimer MODEL}
\onecolumngrid
\subsection*{ Many-body system}

Let us consider $N$ such dimers interacting with each other in a thermal atmosphere at a constant temperature $T$. The similar constituents of different dimers i.e., those with the same functional form of damping, repel each other within some defined proximity and dissimilar constituents attract each other within certain distance keeping in place excluded volume interaction at small distances between them. These dimer are confined to  a circular two-dimensional region. The equations of motion for $N$ such dimer are:

%\begin{widetext}
\begin{eqnarray}\nonumber
\frac{  d{ \mathbf { r}_{1}^{i}  } }{dt} &=& -\frac{1}{\Gamma_{1}( r ^{i})}\nabla_{\mathbf{r}_{1}^{i}} [ V_{h}({r}^i) +  V_{b}({r}_{1}^{i}) ] -\frac{1}{\Gamma_{1}( r ^{i})}\nabla_{\mathbf{r}_{1}^{i}}\sum\limits_{\substack{j\\ i\neq j}} \big[ V_{l}(|\mathbf{r}_{1}^{i} - \mathbf{r}_{1}^{j}|) +   V_{ u }(|\mathbf{ r}_{1}^{i} - \mathbf{r}_{2}^{j}|)\big] + \sqrt\frac{2k_{B}T}{ \Gamma_{1}( r ^{i})} {\boldsymbol{\eta}}_{1}^{i}(t)\\
\frac{  d{ \mathbf { r}_{2}^{i} } }{dt} &=& -\frac{1}{\Gamma_{2}( r ^{i})}\nabla_{\mathbf{r}_{2}^{i}}[V_{h}({r}^i) +  V_{b}({r}_{2}^{i}) ] -\frac{1}{\Gamma_{2}( r ^{i})}\nabla_{\mathbf{r}_{2}^{i}} \sum\limits_{\substack{j\\ i\neq j}} \big[ V_{l}(|\mathbf{r}_{2}^{i} - \mathbf{r}_{2}^{j}|) +   V_{ u }(|\mathbf{ r}_{2}^{i} - \mathbf{r}_{1}^{j}|)\big] + \sqrt\frac{2k_{B}T}{ \Gamma_{2} ( r ^{i})} {\boldsymbol{\eta}}_{2}^{i}(t),
\end{eqnarray}
%\end{widetext}

\onecolumngrid
where $i,j $ $\in$ $\{1,2,\ldots\,N\}$ are dimer indices and  $\mathbf {r }^{i} =\mathbf{r}_{1}^{i}-\mathbf{r}_{2}^{i}$ and $r^i=|\mathbf {r}^i|$. Coordinates $\mathbf{r}_{1}^{i} $ and $\mathbf{r}_{2}^{i}$  are positions of constituent 1 and constituent 2 of the $i$-th dimer where the origin of coordinates is at the centre of the circular boundary. We define $r^i_{1}=|\mathbf {r}^i_{1}|$ and $r^i_{2}=|\mathbf {r}^i_{2}|$ The interaction potential between primary constituents of a dimer (constituent 1 and constituent 2) $V_{h}$ is harmonic. Symbols $\Gamma_{1}(r ^{i})$ and $\Gamma_{2}(r ^{i})$ are coordinate-dependent damping coefficients associated, respectively, to constituent 1  and constituent 2 of $i$-th dimer, which remain similar for all dimers. $\boldsymbol{\eta}_{1}^{i}(t)$ and $\boldsymbol{\eta}_{2}^{i}(t)$  are vector gaussian noises with cartesian coordinate representation as
$\boldsymbol{\eta}_{1}^{i}(t) = {\eta}_{1x}^{i}(t) \hat{{x}} +{\eta}_{1y}^{i}(t) \hat{{y}} $ and $\boldsymbol{\eta}_{2}^{i}(t) = {\eta}_{2x}^{i}(t) \hat{{x}} +{\eta}_{2y}^{i}(t) \hat{{y}} $. Each component represents Gaussian white noise of zero mean and a unit strength. None of these components are cross-correlated i.e. $\langle \eta^i_{ms}\eta^j_{m's^\prime} \rangle = \delta_{ij}\delta_{mm^\prime}\delta_{ss^\prime}$  where $m/m^\prime \in \{1,2\}$ and  $s/s^\prime \in \{x,y\}$.
%$\langle \eta^i_k\eta^j_l \rangle = \delta_{ij}\delta_{kl}$. 
\par
The potential $V_{b}({r}^i_{m})$ is confining particles within circular boundary. The potential $V_{l}(|\mathbf{r}_{m}^{i} - \mathbf{r}_{m}^{j}|)$ is harmonic repulsive in nature maintaining interaction between similar constituents of different dimer and $V_{u}(|\mathbf{ r}_{m}^{i} - \mathbf{r}_{m^\prime}^{j}|)$ is piecewise combination of Lennard-Jones truncated shifted potential and a harmonic repulsive potential to encode the interaction between dissimilar constituents of different dimer. The repulsive part is contributed by harmonic potential and attractive part by truncated shifted Lennard-Jones potential with force and potential going to zero at cutoff $r_{c}$. This kind of choice helps to take into account the repulsive interactions without needing to utilise very short time steps to counter the otherwise steep repulsive landscape of traditional Lennard-Jones potential especially at large well depth. Coordinate dependence of the damping coefficients in simulations is chosen to have following sigmoidal form

\[
\Gamma_{m}( r ^{i}) = \frac{a_{m}}{1+ \exp(-\lambda_{m}(r^{i}-r_{min}))} +b_{m},
\]

\begin{comment}
 \[
\Gamma_{1}( r ^{i}) = \frac{a_{1}}{1+ \exp(-\lambda(r^{i}-r_{min}))} +b_{1},
\]
\[
\Gamma_{2}( r ^{i}) = \frac{a_{2}}{1+ \exp(-\lambda(r^{i}-r_{min}))} +b_{2},
\]
\end{comment}

where $\lambda_{m}$ is the steepness parameter and $a_m$, $b_m$ are constants. The dampings are function of the configuration coordinate of individual dimer and do not depend on inter particle separation of different dimer. This is a simplified approach, however, enough to reveal intended results in the present context. The potentials used  with typical parameter values used in simulation are shown in Fig.1 and have the following structure:$\cdot$

\begin{figure}[htbp]
	\captionsetup{justification=raggedright,singlelinecheck=false}
	\captionsetup{font=footnotesize,labelfont=footnotesize}
	\centering
	   \includegraphics[width=0.70\textwidth]{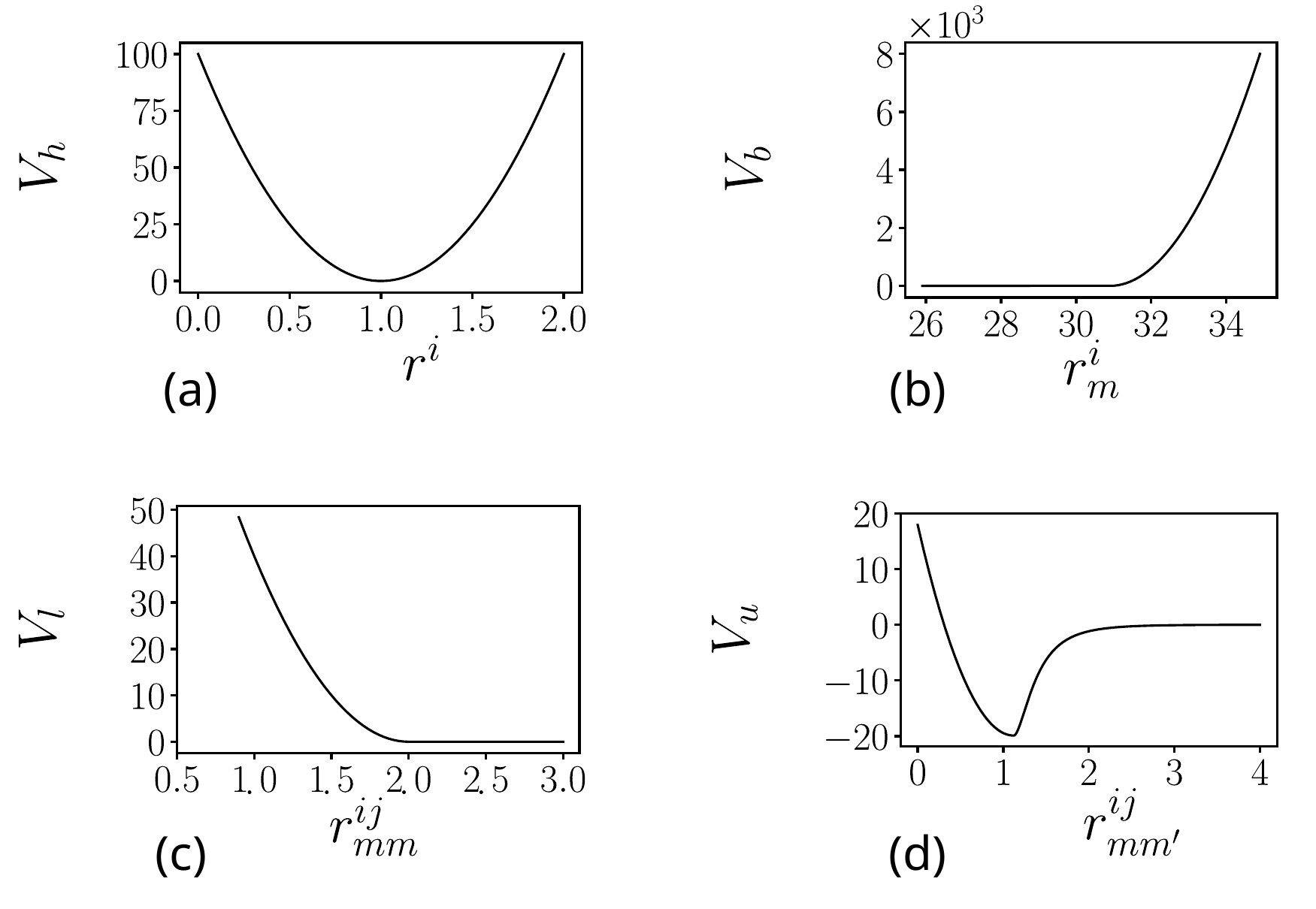}%width =1.0
\caption{Representative profile of potentials: (a) Intra-dimer harmonic potential: $\alpha =200$, $r_{min}=1$. (b) Confining potential: $k_{bo} =1000$, $R = 31$. (c) Similar constituent (like monomer) interaction potential: $\kappa_{l}=80$, $r_{l}=2$. (d) Dissimilar constituent (unlike monomer) interaction potential: $k_{u}= 60$, $\epsilon=20$, $\sigma=1$. }. 
\end{figure}

\begin{equation*}
V_{h}({r}^{i}) = \frac{\alpha}{2} (r^{i} - r_{min})^2,   
\end{equation*}
where $\alpha$ is the spring constant and $r_{min}$ is the equilibrium length for intra-dimer harmonic interaction $V_{h}$.
\begin{equation*}
  V_{b}({r}^{i}_{m}) =
  \begin{cases}
                 0 & \text{if $  r^i_{m} <R,$} \\\\
                \frac{k_{bo}}{2}(r^i_{m}-R)^2  & \text{if $ r^i_{m} \geq R, $}
  \end{cases}
\end{equation*}\\

The constant $k_{bo}$ is repulsion strength of confining potential $V_{b}$, where $R$ is the radius of circular boundary and  $r^i_{m}=|\mathbf {r}^i_{m}|$. Potential $V_{l}$ that corresponds to repulsive interaction between the like (similar) constituents of different dimers has the structure
\begin{equation*}
%\[
  V_{l}(\mathbf{r}^{ij}_{mm}) =
  \begin{cases}
                 \frac{1}{2}\kappa_{l}(r^{ij}_{mm}-r_{l})^2& \text{if $r^{ij}_{mm}<r_{l},$} \\\\
                0 & \text{if $r^{ij}_{mm}\geq r_{l},$}
  \end{cases}
%\]
\end{equation*}
where, $\mathbf{r}^{ij}_{mm} =\mathbf{ r}_{m}^{i} - \mathbf{r}_{m}^{j}$ and $r^{ij}_{mm} =|\mathbf{r}^{ij}_{mm}|$ is the magnitude of separation between same constituent (same constituent/monomer index $m$) of different dimers (different dimer indices $i$ and $j$), with the repulsion strength $\kappa_{l}$. Constant $r_{l}$ is the distance below which the same constituents start feeling repulsive force. 

\par
The interaction potential $V_{u}$, which is piecewise combination of two potentials such that the repulsive part is governed by shifted harmonic repulsion and attractive part by a typical L-J truncated shifted potential is defined as  

\[
  V_{u}(\mathbf{r}^{ij}_{mm'}) =
  \begin{cases}
                 (\Tilde{\Phi}_{u}-V^{*}_{LJ})+ F^{*}_{LJ}\times(r^{ij}_{mm'}-r_{c})&\text{if $r^{ij}_{mm'}\leq r^{'},$} \\ \\
  
                 (V_{LJ}-V^{*}_{LJ})+ F^{*}_{LJ}\times(r^{ij}_{mm'}-r_{c})&\text{if $r^{'}\leq r^{ij}_{mm'}\leq r_{c},$} \\ \\
                0 & \text{if $r^{ij}_{mm'} \geq r_{c}.$}
  \end{cases}
\]

 where, $\mathbf{r}^{ij}_{mm'} =\mathbf{ r}_{m}^{i} - \mathbf{r}_{m'}^{j}$ and $r^{ij}_{mm'}=|\mathbf{r}^{ij}_{mm'}|$
 is the magnitude of separation between unlike constituents. The potential $V_{u}$ is cut off at distance $r_{c}$, as one reaches $r_{c}$ from smaller separations, the force and potential smoothly go to zero and above it continue to remain zero. The standard L-J potential is defined as $V_{LJ}(r^{ij}_{mm'}) = 4\epsilon[(\sigma/r^{ij}_{mm'})^{12}-(\sigma/r^{ij}_{mm'})^{6}] $, where $\epsilon$ is the depth of well, $\sigma$ is a length parameter and $r^{'} = 2^{\frac{1}{6}}\sigma$ is the value of position at minimum of a standard L-J potential. $F_{LJ}$ is the corresponding L-J force. The $\Tilde{\Phi}_{u}$ is the shifted harmonic repulsion defined as $\Tilde{\Phi}_{u}(r^{ij}_{mm'}) = \frac{k_{u}}{2}(r^{ij}_{mm'}-r')^2 - \epsilon$, with $k_{u}$ measures of strength of harmonic repulsion. Constants $V^{*}_{LJ}$ and $F^{*}_{LJ}$ are the values of standard Lennard-Jones potential and force at cutoff distance $r_{c}$  respectively, with $r_{c}$ set to be $4 \sigma$ in simulations. $V_{u}$ and the corresponding force is continuous at $r^{'}$ by construction. 

\onecolumngrid
\subsection*{Simulation methods}
The dynamics (Eq.(6)) is transformed in position coordinate to $4N$ coupled Brownian equations in cartesian system of coordinates for our simulation. Dimers are initialized in random configuration and Euler-Maruyama algorithm is implemented to simulate these $4N$ coupled Brownian dynamics equations. The following updating scheme is used for our simulations:

\begin{enumerate}
\item[(1)]Assign random positions to dimers at beginning of the simulation.
\item[(2)]Corresponding to the positions, calculate the respective forces.
\item[(3)] Update the position of each dimer using the forces calculated in step (2) by employing explicit Euler-Maruyama discretization scheme.
\item[(4)]  Go back to step (2) and repeat.
\end{enumerate}

\par
Typical parameters corresponding to coordinate-dependent damping used in simulations that follow, for monomer 1 and monomer 2 (unless stated otherwise) are $a_{1}=-0.3$, $b_{1}=0.4$ and $\lambda_{1}=100$; and $a_{2}=-0.1$, $b_{2}=0.3$ and $\lambda_{2}=100$ respectively. Typical structures of the potentials are shown in Fig.1 for a particular set of parameters. Simulations are run for every set of parameters of interaction always under symmetry-broken ($\Gamma_1(r^i)\neq\Gamma_2(r^i)$) and symmetry-unbroken ($\Gamma_1(r^i)=\Gamma_2(r^i)$) conditions. This is done to compare directed transport in the symmetry-broken case with that in symmetry-unbroken case which does not show any average directed motion in the similarly formed many-body structures. Link of the movies of pair of such symmetry-broken and unbroken situations under exactly the same other interaction parameters are given in the respective figure captions of Fig.2. 
\par
All simulations use a new random seed each time step and it has been checked that qualitative results remain the same for given set of parameters independent of the random seed. We have set $k_{B}T$ and $r_{min}$ as our reference for energy and distance scale respectively and both have been set to unity. The number of dimers have been fixed to thirty in all simulations. Videos have been produced by harnessing matplotlib library, using the data of positions of dimers for both symmetry broken case and a symmetry-unbroken case. In case the videos generated are long or too slow, those have then been sped up by using a video editor.
\par
Several symmetry unbroken situations at a wide range of parameter values are checked to observe no average directed motion. Parameters of the functions of the symmetry-broken cases mentioned in this paper are for the best representation, however, it has been checked that, results over wider similar parameter ranges remain qualitatively same. The average velocity under symmetry-broken scenario would obviously depend on the value of the parameters in the damping function, but, there always are some directed motions under the symmetry-broken conditions even when the broken symmetry is marginal. Some phase diagrams for the continuous emergence of directed motion under a growth of broken symmetry is also shown in what follows. 

\section{Results}
\subsection*{ Characteristic motion of various structures}

\begin{figure}[htbp]
	\captionsetup{justification=raggedright,singlelinecheck=false}
	\captionsetup{font=footnotesize,labelfont=footnotesize}
	\centering
	   \includegraphics[width=1.0\textwidth]{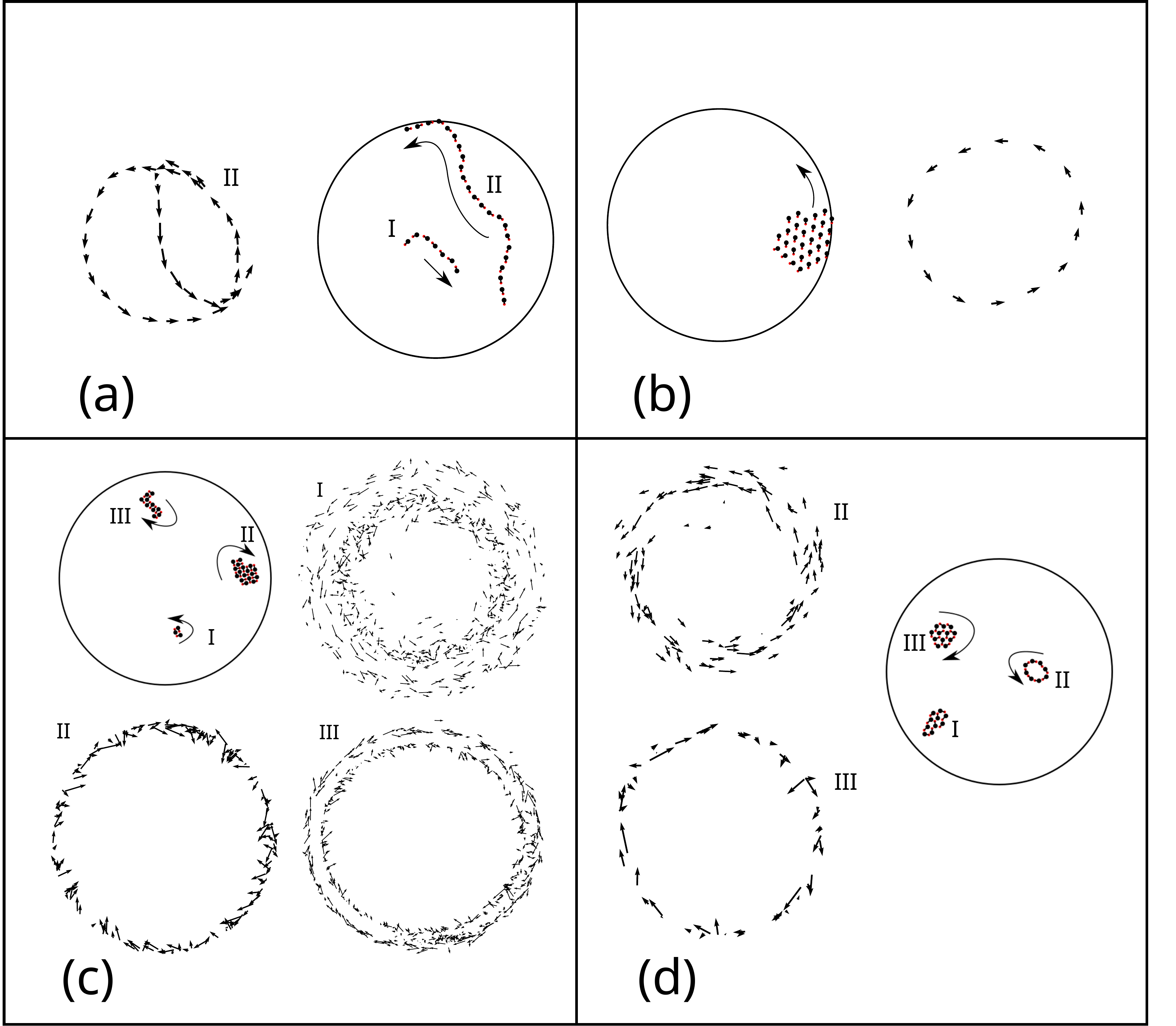}

\caption{Parameters : 2(a): $\sigma=1.5$, $r_{l}=3$ and $\Delta t =10^{-4}$ and 2(b): $\sigma=2$, $r_{l}=3$ and $\Delta t =10^{-5}$. The movie of the time evolution of configurations in 2(a) and 2(b) sped up by ten times and five times respectively  can be found here \href{https://www.youtube.com/watch?v=raPftgQ-RPU}{ 2(a)} and \href{https://www.youtube.com/watch?v=_6jkxK9bgRo&t=142s}{2(b) }; 2(c):  $\sigma=1$, $r_{l}=1.8$ and $\Delta t =10^{-5}$ and 2(d): $\sigma=1$, $r_{l}=2.2$ and $\Delta t =10^{-4}$. The movie of the time evolution of structures in 2(c) and 2(d) sped up by ten times and one and half times respectively can be found here \href{https://www.youtube.com/watch?v=GFWTYNqKp5s} {2(c)} and \href{https://www.youtube.com/watch?v=CThM4UdfJKM&t=81s}{2(d)}.
}

\end{figure}
In Fig.2, we show snapshots of phases under different interaction between dimers. Since these phases are dynamic, we also provide links to the movies showing dynamics of these phases in respective sub-figure caption. Parameters are mentioned in the caption of sub-figures as well. For the phase shown in Fig.2(a), we ran simulation for $5\times 10^{6}$ iterations and recorded positions of dimers every 100 iterations. A step of iteration for all the simulations includes update of position of all the particles by one time step ($\Delta t$). All the sub-figures in Fig.2 comprise of parts, where the figure in the circle gives a snapshot of the dimer configurations, other vector-plots show velocity of these structures on average over a given time interval. Linear queued chains of dimers were formed as shown in Fig.2(a). The chain-like flocks of dimer generally display translational motion unless they form a ring like structure which rotates in a particular direction dictated by the broken symmetry of the individual dimers. 

\par
Fig.2(b) is for a different interaction regime. We show results of simulation for $5\times10^7$ iterations where the positions of the dimers are recorded at every $1000$ iterations. Initially small cyclic chains of associated tri-dimer and quad-dimer are formed. The intermediate structures formed by the association of the three and four dimers when exist in a closed loop always rotate in a particular direction. The direction of the rotation is selected by the direction of motion of individual dimers. At large times, many-body interactions in this parameter regime leads to parallel aligned dimers akin to ferromagnetic order. This eventually forms a single cluster that swarms through on the confined plane. 

\par
Fig.2(c) shows snapshot of results of simulation we ran for $5 \times 10^{7}$ iterations and positions recorded every $ 1000$ iterations. At small times, dimers form  chiral-symmetry broken tri-dimer and quad-dimer which then combine and form a large chiral-symmetry broken rectangular lattice like structures which rotate depending upon the broken symmetry following the direction of motion of the individual dimers. In Fig.2(c) and Fig.2(d), we display the local time averaged velocity vector of one member of each cluster depicted by arrow with the tail of arrow being fixed at local time averaged position of the member as seen from centre of masses of respective cluster. Three members form a cluster in I as shown in Fig.2(c), they are less compactly arranged, therefore, fluctuations are more. The cluster II is more dense and thus strength of binding forces is more than thermal fluctuation responsible for displacing dimers considerably from their mean position with respect to centre of mass of cluster. This is responsible for the velocity-plot being less fluctuating. The cluster III  undergoes rearrangement in it's structure depicted by transition of arrows from one phase line to other and the less fluctuation in velocity plot is obviously attributed to compactness of structure.     
   
\par
To see a smoothed velocity plot, a local average over velocity and position over time duration equivalent to 200 subsequent recorded positions is performed for cluster II. The local (small time) average angular velocity on calculation came out to be about 0.2 per step of iteration in clockwise direction. We checked that the average angular velocity and qualitative nature of phase plot doesn't change appreciably by average obtained over time duration corresponding to 100, 150, 250, 300 subsequent recorded positions. Similar scheme is followed for plotting phase plot of other clusters of Fig.2(c). The steady average value of angular velocity of each cluster in 2(c) and 2(d) is evaluated by first doing a long-time average over time series of angular velocity of each member of cluster, and then doing an arithmetic average over all the particles constituting a cluster. For obvious reasons, this averaging is done on clusters when no particle entered or left the cluster and number of particles in cluster remain fixed. The steady state average angular velocity of clusters I, II and III are found to be 4.6, 0.2 and 0.59 per step of iteration respectively with averaged direction of rotation indicated by arrows in vector plot figure. 

\par
In Fig.2(d), we show snapshot of results for simulation of $10^{7}$ iterations, with positions being recorded every $ 500$ iterations. The dimers arrange themselves in triangular lattice like structures that shows uni-directional rotation. 
Cluster II undergoes rearrangement by forming contracting or expanding dimer ring like structure that rotates in one direction, and this is why the velocity plot shows spread. Cluster III on the other hand forms a more compact structure with uni-directional rotation, thus, the phase plot is less fluctuating. 
For smoothing out the trajectories of cluster II, a local time averaged velocity and position over time corresponding to subsequent 10 positions is taken into account. The average angular velocity on calculation came out to be about 2.8 per step of iteration in anti-clockwise direction which does not change appreciably by doing a local overage over 5 or 15 subsequent position instead of 10. Similar scheme is followed for other clusters. The steady state average value of angular velocity of cluster I, II and III in 2(d) were found out to be 0, 2.8 and 0.38 per step of iteration. 

\subsection*{Emergence of motion from broken symmetry}

Directed transport of these collection of dimers under equilibrium conditions arises due to the broken symmetry as has been defined. To relate the broken symmetry to the emergence of directed motion in these many body systems we need to develop an order parameter that captures the characteristic dynamics so that a quantitative idea of emergence of motion could be developed. To this goal, we use two different order parameters where the former is for the phases depicted in Fig.2(a) and (b) and the latter will correspond to the rest of the cases where the motion is predominantly rotational. Plots of these order parameters (Fig.4 and Fig.5) against the parameter that quantifies broken symmetry clearly shows emergence of transport with the breaking of the symmetry by damping.

We see that a cluster of symmetry-broken dimers show net transport in the system, whereas a cluster of symmetry-unbroken dimers show random jiggling. Now, the net transport can be translational, or rotational or a mixture of both. Inspired from \cite{alaimo2016microscopic,lober2015collisions,hiraoka2017collective}, a meaningful order parameter $\phi$ is defined to quantify a net translation motion along circular boundary for the structures formed in the parameter regimes of Fig.2(a) and (b). We defined an instantaneous measure $\phi(t_{k})$ that measures relative directional alignment and motion of group of dimers with respect to circular boundary of a cluster at time $t_{k}$. 

\[
\phi(t_{k}) = \frac{1}{N^{*}}\Big| \sum_{i=1}^{N^{*}} \frac{ \boldsymbol{v}_{i}(t_{k})\cdot\hat{{e}}_{\theta_{i}}(t_{k})}{ |{\boldsymbol{v}_{i}}(t_{k}) | } \Big|,
\]
where, $N^{*}$ denotes total numbers of dimers constituting a cluster at time $t_{k}$.
Velocity $\boldsymbol{v}_{i}(t_{k})$ is that of the centre of mass of the $i$-th dimer at time $t_{k}$. Angle $\theta_{i}(t_{k})$ is the polar angle made by dimer's centre of mass to $x$-axis at time $t_{k}$, $\hat{{e}}_{\theta_{i}}(t_{k}) = -\sin \theta_{i}(t_{k}) \hat{{x}} +  \cos \theta_{i}(t_{k}) \hat{{y}}$ is the instantaneous  tangential unit vector corresponding to $i$-th dimer. $\phi(t_{k}) =1$ would imply that all dimers of a cluster are aligned and moving in same tangential direction along inner circumference of circle at a time instant $t_{k}$. A schematic description of the quantities used in $\phi(t_{k})$ are shown in Fig.4(a). A time series of $\phi(t_{k})$ is generated. We divide the time series into many small parts and calculate a local time averaged quantity called $\phi_{M}$. 
\[
\phi_{M} = \frac{1}{M} \sum_{k=k_{1}}^{k_{1}+M}\phi(t_{k}),
\]
where, $k_{1}$ denotes time index at beginning of small time series, $k_{1}+M$ denotes time index at ending of small time series and $M$ is number of points in time interval over which local time average is obtained.

An average over all such $\phi_{M}$'s gives us the order parameter $\phi$.
\[
\phi = \langle  \phi_{M} \rangle = \frac{1}{N'} \sum_{j'=1}^{N'}(\phi_{M})^{j'}.
\]
where, $N^{'}$ denotes the total number of $\phi_{M}$'s over which average is obtained and $j^{'}$ is a counter index to keep track of number of $\phi_{M}$'s.

\begin{figure}[htbp]
	\captionsetup{justification=raggedright,singlelinecheck=false}
	\captionsetup{font=footnotesize,labelfont=footnotesize}
	\centering
	   \includegraphics[width=1.0\textwidth]{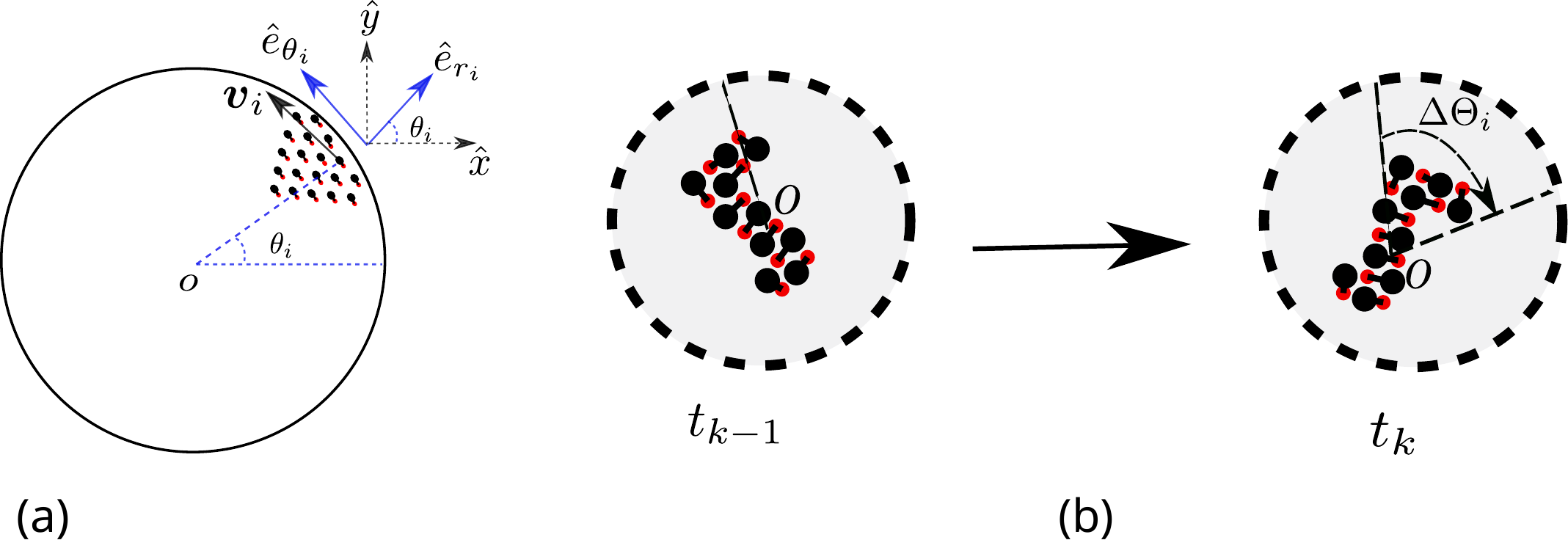}

\caption{(a) Polar angle $\theta_{i}$, it's corresponding polar unit vectors and velocity $\boldsymbol{v}_{i}$ shown for the  $i$-th dimer. (b) Polar angle $\Delta\Theta_{i}$ swept by $i$-th dimer of a cluster, with respect to centre of mass of cluster O as time elapses from  $t_{k-1}$ to $t_{k} $. The instantaneous angular velocity of $i$-th dimer calculated as: $\omega_{i}(t_{k}) = \frac{\Delta \Theta_{i}}{t_{k}-t_{k-1}}$ }.

\end{figure}

In all the cases to quantify motion using order parameter $\phi$ and $\Omega$, we fix $a_{1}=a_{2}=-0.3$, $b_{1}=b_{2}=0.4$ and $r_{min}=1$ for both the monomers of all the dimer. We also set $\lambda_{1}=100$ for monomer $1$ of all dimers and systematically vary $\lambda_{2}$ from $0$ to $100$ for monomer $2$. A symmetry unbroken case here corresponds to $\lambda_{1}=\lambda_{2}=100$ and we expect no net transport in such a scenario.  

\par
In order to quantify overall local uni-directional rotational motion of group of dimers forming a stable cluster, we firstly define instantaneous angular speed  $\omega_{i}(t_{k})$  at time $t_{k}$ associated with each member of a cluster as seen from it's centre of mass, as the polar angle $\Delta \Theta_{i}$ traced by $i$-{th} dimer as we go successively from time instant $t_{k-1}$ to $t_{k}$. 

\[
\omega_{i}(t_{k}) = \frac{\Delta \Theta_{i}}{t_{k}-t_{k-1}} 
\]

 A time averaged angular speed $\Bar{\omega}_{i}$ is then evaluated from the time series of angular speed for each member of the cluster.
\[
\Bar{\omega}_{i} = \frac{1}{N'}\Big| \sum_{k=k_{2}}^{k_{2}+N'}  \omega_{i}(t_{k})  \Big|,
\]
where, $k_{2}$ denotes time index at beginning of time series, $k_{2}+N'$ denotes time index at the ending of the same over which time average is obtained and $N^{'}$ denotes total number of data points used to calculate the average. 

The steady rate of rotation of cluster $\Omega$ is then obtained by doing an arithmetic average over time averaged valued angular speed $\Bar{\omega}_{i}$ of members of the cluster.

\[
\Omega =\langle  \Bar{\omega}_{i} \rangle = \frac{1}{N^{''}} \sum_{i=1}^{ N^{''}}\Bar{\omega}_{i}
\]
where, $N^{''}$ denotes the the total number of members in a cluster.
Even for the same $\lambda_{2}$ and other fixed parameters, different clusters would have different angular velocities depending on the arrangement of dimers. It becomes clear from Fig.4 and Fig.5 that as one reaches from symmetry-broken to symmetry-unbroken case, the net transport ceases to exist. 

\begin{figure}[htbp]
	\captionsetup{justification=raggedright,singlelinecheck=false}
	\captionsetup{font=footnotesize,labelfont=footnotesize}
	\centering
	   \includegraphics[width=0.9\textwidth]{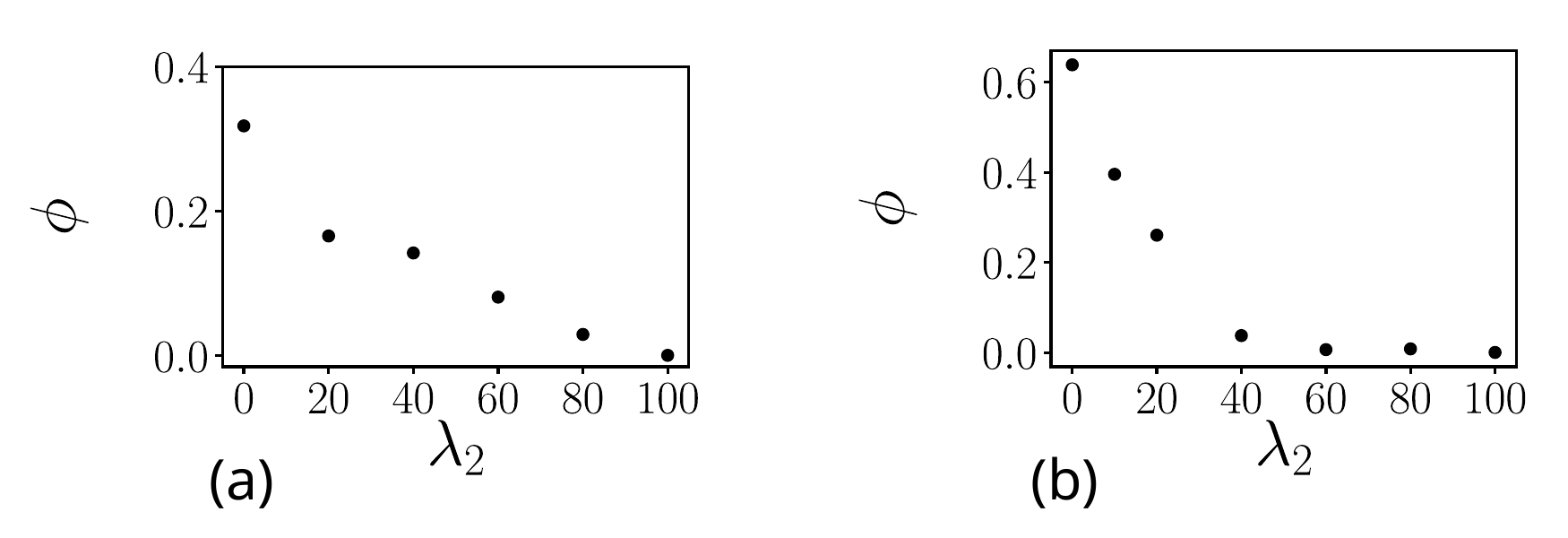}
    
\caption{Variation of order parameter $\phi$ with $\lambda_{2}$, keeping other damping parameters fixed: $a_{1}=a_{2}=-0.3$, $b_{1}=b_{2}=0.4$ and  $\lambda_{1} =100$. The interaction parameters for Fig.4(a) and 4(b) are for the exact same regimes that correspond to kind of motion shown in 2(a) and 2(b) respectively. 
}

\end{figure}

\begin{figure}[htbp]
	\captionsetup{justification=raggedright,singlelinecheck=false}
	\captionsetup{font=footnotesize,labelfont=footnotesize}
	\centering
	   \includegraphics[width=0.9\textwidth]{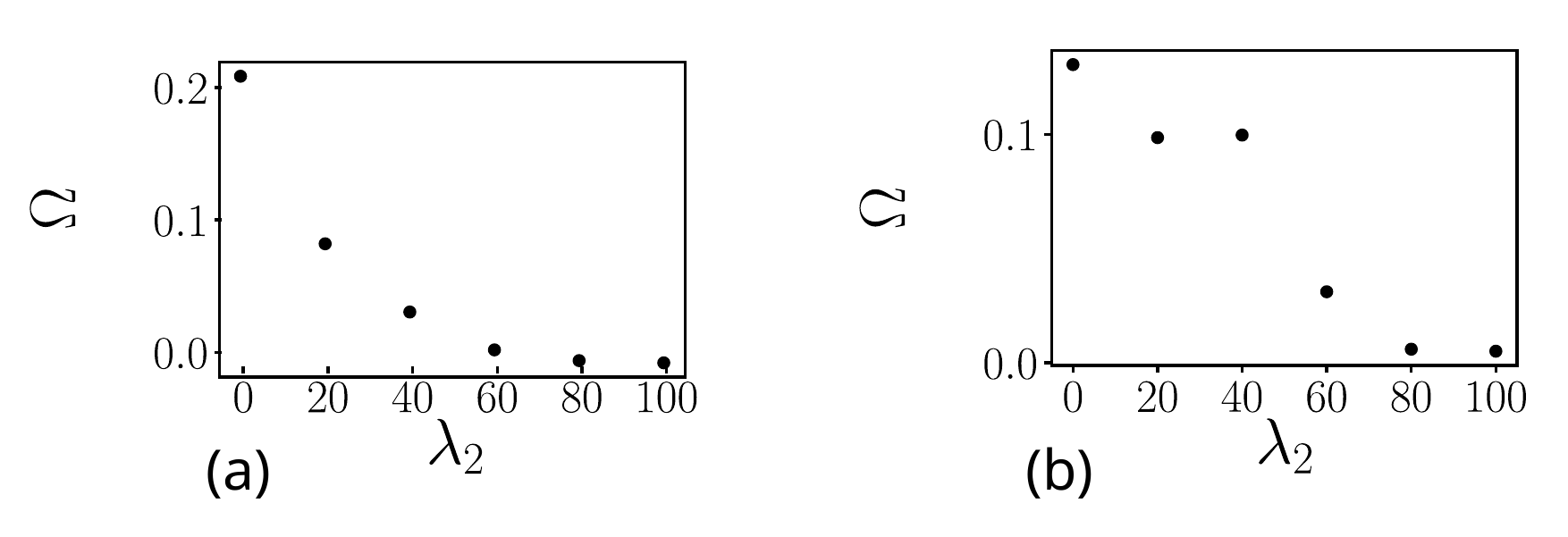}
	   
\caption{Variation of order parameter $\Omega$ with $\lambda_{2}$, keeping other damping parameters fixed : $a_{1}=a_{2}=-0.3$, $b_{1}=b_{2}=0.4$ and  $\lambda_{1} =100$. The interaction parameters for Fig.5(a) and 5(b) are for the exact same regimes that correspond to kind of motion shown in 2(c)  and 2(d) respectively.
}

\end{figure}

\section*{Discussion}

In this paper we have explored the thermal directed motion of various structures formed by interacting dimers whose constituent particles have a configuration dependent damping and diffusion related by the Stokes-Einstein (fluctuation-dissipation) relation. This configuration dependent damping/diffusion in a heat-bath basically drives the motion in the absence of any active force. The essential condition for the existence of such motions under symmetry broken by damping is that - run the dynamics in accordance with It\^o-convention - i.e. do not introduce any anticipating (correlated) noise as is done in Stratonovich or Stratonovich-like conventions. This means, while numerically evolving the given dynamics (Eq.6), at every time step ($\Delta t$) of evolution, we take the noise strength (diffusivity) to be determined by where the particle is sitting at the start of the interval $\Delta t$. This is the simplest way to implement It\^o-convention which perfectly corresponds to a thermal noise because it does not introduce any correlation.
\par
The results which have been first predicted in \cite{bhattacharyay2012directed} and later used in \cite{sharma2020conversion} in one-dimensional case are extended here to higher dimensional space being assisted by many-body interactions. This many-body effect is remarkable in relation to the fact that it has never been envisaged under equilibrium conditions and one in general employs non-equilibrium drive (active systems) to achieve such effects. The non-equilibrium drive practically introduces correlations in the noise, however, in It\^o-convention, the noise is completely correlation free. What makes the difference here, to what has been believed for so long, is the functional difference of the It\^o-distribution from the Boltzmann-distribution. Where the It\^o-distribution does sustain the broken symmetry by making the average centre of mass velocity non-zero while keeping the configuration velocity zero, this is not in general possible by Boltzmann distribution. Forcing of the Boltzmann distribution in such cases (under coordinate dependent damping/diffusion) has so far eluded one realizing the general possibility of directed transport in equilibrium although coordinate dependence of damping is known for almost over a century and the It\^o-convention appeared in literature in the year 1944.

\par
The very existence of the It\^o-distribution replacing the Boltzmann one modifies the free energy surface. Even on top of that, the most striking consequence of existence of such motions under It\^o-distribution is that the pathway of structure formation in complex molecules or collection of them is not only determined by fluctuations over a free energy surface, rather, is complemented by characteristic motions at different free energy states. Uncorrelated thermal fluctuations resulting in structure dependent characteristic motion is a general outcome under coordinate dependence of damping and diffusion. Therefore, a Monte Carlo simulation in such cases is not enough to mimic the reality and one needs to explicitly take into account the dynamics. In our view, this new phenomenon needs some serious attention at present and probably holds a general key to help us understand many complex processes including those in living systems. 

 %\clearpage
 
\begin{acknowledgments}
Mayank would like to acknowledge Abhishek Anand for his help and inputs in creating video animation. Mayank would also like to acknowledge Ritam Pal for useful discussions.
\end{acknowledgments}

\nocite{*}
\bibliography{biblographysample}% Produces the bibliography via BibTeX.

\newpage
\onecolumngrid

 \end{document}